# Code-free development and deployment of deep segmentation models for digital pathology

**Henrik Sahlin Pettersen**[*,1,2,3], **Ilya Belevich**[4], **Elin Synnøve Røyset**[1,2,3], **Erik Smistad**[5,6], **Eija Jokitalo**[4], **Ingerid Reinertsen**[5,6], **Ingunn Bakke**[2,3], and **André Pedersen**[2,5]

1. Department of Pathology, St. Olavs hospital, Trondheim University Hospital, Trondheim, Norway.

2. Department of Clinical and Molecular Medicine, Faculty of Medicine and Health Sciences, NTNU - Norwegian University of Science and Technology, Trondheim, Norway.

3. Clinic of Laboratory Medicine, St. Olavs hospital, Trondheim University Hospital, Trondheim, Norway.

4. Electron Microscopy Unit, Institute of Biotechnology, Helsinki Institute of Life Science, University of Helsinki, Helsinki, Finland.

5. Department of Health Research, SINTEF Digital, Trondheim, Norway.

6. Department of Circulation and Medical Imaging, Faculty of Medicine and Health Sciences, NTNU - Norwegian University of Science and Technology, Trondheim, Norway.

**\* Correspondence:**

Henrik Sahlin Pettersen

henrik.s.pettersen@ntnu.no

**Abstract**

Application of deep learning on histopathological whole slide images (WSIs) holds promise of improving diagnostic efficiency and reproducibility but is largely dependent on the ability to write computer code or purchase commercial solutions. We present a code-free pipeline utilizing free-to-use, open-source software (QuPath, DeepMIB, and FastPathology) for creating and deploying deep learning-based segmentation models for computational pathology. We demonstrate the pipeline on a use case of separating epithelium from stroma in colonic mucosa. A dataset of 251 annotated WSIs, comprising 140 hematoxylin-eosin (HE)-stained and 111 CD3 immunostained colon biopsy WSIs, were developed through active learning using the pipeline. On a hold-out test set of 36 HE and 21 CD3-stained WSIs a mean intersection over union score of 96.6% and 95.3% was achieved on epithelium segmentation. We demonstrate pathologist-level segmentation accuracy and clinical acceptable runtime performance and show that pathologists without programming experience can create near state-of-the-art segmentation solutions for histopathological WSIs using only free-to-use software. The study further demonstrates the strength of open-source solutions in its ability to create generalizable, open pipelines, of which trained models and predictions can seamlessly be exported in open formats and thereby used in external solutions. All scripts, trained models, a video tutorial, and the full dataset of 251 WSIs with ~31k epithelium annotations are made openly available at https://github.com/andreped/NoCodeSeg to accelerate research in the field.

# 1    Introduction

Visual evaluation of histopathological whole slide images (WSIs) is the gold standard for diagnosing an array of medical conditions ranging from cancer subtyping and staging to inflammatory and infectious diseases. The increasing shortage of pathologists in combination with continually increasing biopsy load and the increasingly evident lack of reproducibility of diagnoses between pathologists call for the application of novel methods to improve both diagnostic efficiency and reproducibility (van der Laak et al., 2021). Application of deep learning-based methods to histopathological WSIs holds promise of improving diagnostic efficiency and reproducibility, but is largely dependent on the ability to write computer code or buy commercial solutions. The introduction of large-scale digitization of histopathological WSIs has moved several pathology departments away from manual microscopy diagnostics to diagnosing digitized WSIs on computer screens (Jahn et al., 2020). The successful application of deep learning-based classification and segmentation of WSIs holds great promise for a continually increasing introduction of computer assisted diagnostics for pathologists, possibly alleviating both pathologist workload and increasing reproducibility (Djuric et al., 2017; Srinidhi et al., 2021). Many current solutions are either commercial software with limited transparency of the applied algorithms, limited export/import capability for other software, and limited availability for diagnostic departments with strained budgets. Existing commercial solutions include software such as Visiopharm[1], Halo AI[2], and Aiforia[3], but also, open-source alternatives such as MONAI-Label[4], H-AI-L, QuickAnnotator (Lara et al., 2021; Lutnick et al., 2019; Miao et al., 2021; von Chamier et al., 2021), and ZeroCostDL4Mic (von Chamier et al., 2021). These open-source solutions, however, either lack a full annotation, training and visualization pipeline, require some degree of programming experience, or use commercial servers. This calls for the development and use of open-source solutions that enable transparency of the image analysis pipelines, the possibility of exporting and importing results and data between applications and use of local data without the requirement of uploading restricted images to commercial serves.

The open-source software QuPath is a user-friendly solution for WSI analysis (Bankhead et al., 2017). Its tools offer means for tumor identification and biomarker evaluation using conventional non-deep learning-based machine learning methods with possibilities of batch-processing and scripting, as well as communication with auxiliary image analysis applications such as ImageJ. However, no deep learning-based image segmentation functionality exists for QuPath to date, except for the possibility of calling the StarDist nucleus segmentation method (Schmidt et al., 2018) from a customizable script in the latest 0.3.x release.

---

[1] https://visiopharm.com/visiopharm-digital-image-analysis-software-features/ai-features/),
[2] https://indicalab.com/halo-ai/
[3] https://www.aiforia.com/
[4] https://github.com/Project-MONAI/MONAILabel/

Application of deep learning approaches to biological imaging during recent years has significantly boosted our capabilities to segment structures of interest from collected images and make them ready for visualization and quantitative analysis (Moen et al., 2019). Despite the potential quality of generated results, use of deep learning in routine research projects is still quite limited. This limitation is mostly due to a relatively high threshold barrier that is hard to overcome by researchers without extensive knowledge of computer science and programming experience. The typical deep learning workflows require knowledge of deep learning architectures, Python programming abilities, and general experience with multiple software installations. The code-free solution DeepMIB was published to help with all these aspects and with a hope to make deep learning available to a wider community of biological researchers (Belevich & Jokitalo, 2021). DeepMIB is a user-friendly software package that was designed to provide a smooth experience for training of convolutional neural networks (CNN) for segmentation of light and electron microscopy datasets. It is suitable for anyone with only very basic knowledge of deep learning and does not require computer programming skills.

DeepMIB comes bundled with Microscopy Image Browser (Belevich et al., 2016), which is a free, open-source software package for image processing, segmentation, and quantification of microscopy datasets. Both packages are written with MATLAB, they are easy to install, and can be used either under the MATLAB environment or as a stand-alone application on Windows, macOS, or Linux.

Image segmentation in DeepMIB is organized as a step-by-step workflow, which starts with selection of a CNN architecture (2D or 3D, U-Net (Ronneberger et al., 2015) or SegNet (Badrinarayanan et al., 2017)), and definition of the most central training hyperparameters. The provided architectures are efficient and are shown to generate generalizable models even with sparse training data (Falk et al., 2019). To extend the training base, DeepMIB comes with multiple (19 for 2D, and five for 3D) augmentation filters that can be individually configured, previewed, and tuned to fulfill the needs of a specific project. The resulting CNN models can be used to predict images directly in DeepMIB or be exported to ONNX format. DeepMIB further provides the ability to test the performance of the trained model on an unseen test set with ground truth labels and evaluate the network performance using multiple metrics, such as accuracy (ACC), dice similarity coefficient (DSC) and intersection over union (IoU). The MIB software is openly available on GitHub[5].

Multiple studies propose deep learning solutions for computational pathology (Srinidhi et al., 2021). However, only some make their trained models openly available; even if they were, using them generally requires programming experience. In digital pathology, this is especially challenging due to the large image sizes of up to 200,000 x 100,000 color pixels, which makes it computationally demanding to deploy models and visualize the predictions with the WSI. Although MIB is able to run inference on the WSI level, the browser is not suitable for displaying such large images, only supports semantic segmentation models, and does not have a streamlined algorithm to exclude prediction on background glass areas. This slows viewing speed, versatility, and prediction runtime. FastPathology (Pedersen et al., 2021) was developed to offer a user friendly direct WSI prediction viewer to pathologists. The software is free, open-source, and focused on high-performance computing to minimize memory usage and runtime. The software is based on the C++ library FAST (Smistad et al., 2015; Smistad et al., 2019).

---

[5] https://github.com/Ajaxels/MIB2

FastPathology enables the user to deploy deep learning methods directly from the Graphical User Interface (GUI). The software includes a rapid, pyramidal viewer for visualizing WSIs and supports overlays of segmentations. New models can be imported without implementation, by defining a FAST text pipeline that contains information about the model and how it should be handled. The software supports various inference engines i.e., TensorRT, OpenVINO, and TensorFlow (Abadi et al., 2016). TensorRT enables the fastest graphical processing unit (GPU) inference, whereas OpenVINO is among the fastest central processing unit (CPU) alternatives. The recommended format is ONNX, as both OpenVINO and TensorRT support it. FastPathology is openly available on GitHub[6], including trained models and test data.

Here we present a pipeline for developing and deploying high performance deep segmentation models for WSIs using three software packages, each specialized in different parts of the workflow (Figure 1). QuPath is efficient for quick annotations of WSIs, DeepMIB provides capabilities for training CNNs without programming, and FastPathology for efficient inference and visualization of full resolution model predictions with the WSI. The proposed pipeline is demonstrated on a use case of segmentation of colon epithelium and is shown to produce models that perform at a clinical acceptable accuracy and runtime level.

## 1.2  Example application

The human gut mucosa comprises both non-immune and immune cells working together in a complex manner to maintain mucosal immunity. In lamina propria there are a broad range of different innate and adaptive immune cell subtypes that are separated from gut content and microbiota by a single layer of intestinal epithelial cells at the surface. These specialized epithelial cells have a pivotal role in producing mucus and antimicrobial factors, or immunomodulating cytokines involved in crosstalk between the different systems, in addition to being a physical barrier. Intermingled between the epithelial cells resides a population of intraepithelial T lymphocytes (IEL), many of which are unconventional T cells. These cells have the characteristics of both innate and adaptive immunity, and they can move and surveil the epithelium. This makes them able to respond rapidly and diverse as an effective first line defense against microbe invasion in addition to being important for maintenance of mucosal homeostasis (Lutter et al., 2018; Olivares-Villagomez & Van Kaer, 2018). Dysregulation of IELs is generally correlated to loss of mucosal barrier integrity and is implicated in the pathogenesis of several gut disorders like infections with bacteria, parasites and viruses, inflammatory processes like inflammatory bowel disease, lymphocytic colitis, and celiac disease, and possibly also tumor development (Lutter et al., 2018). A lot is still unknown about the functions and clinical significance of the different IEL subtypes, and more research is needed (Hu & Edelblum, 2017; Lutter et al., 2018). Tools that provide objective and reproducible quantitative data from tissue sections will open new doors in research and allow for new questions to be posed.

---

[6] https://github.com/AICAN-Research/FAST-Pathology

For inflammatory disorders of the GI tract that involves numerical definitions, like celiac disease or lymphocytic colitis, quantification of IEL is part of the pathologist's job. This can be done by roughly giving a visual estimate, or by manual counting of smaller areas and then make a global estimate based on that. Looking at tissue sections, the eyes are more easily drawn to the areas with the highest densities and could possibly lead to an overestimation of the number of IELs. A tool for epithelium segmentation that enables automated quantification of IELs could serve as a calibration instrument for pathologists. It can save pathologists from spending time and energy on something that can be done much more objectively by a machine. It can be of great value in research on the epithelial immune microenvironment in inflammatory and neoplastic disorders. In the present study, we have included both HE and CD3 stained images to demonstrate the potential use of this technique both for quantifying different populations of intraepithelial immune cells with the help of immunostaining and the potential (by further annotation and training) to quantify e.g., intraepithelial granulocytes directly on HE stained images. Quantification of CD3 immunostained IELs after epithelial segmentation can be achieved with high accuracy in QuPath, but is not demonstrated as part of this publication. Further developing deep learning-based models for segmentation of other important mucosal structures (e.g., lymphoid aggregates, basal plasmacytosis, specific cell types, tumors), and for other types of immunohistochemical evaluations to integrate information of protein expression, cell types and tissue structure, would vastly expand the value of this tool for research and in diagnostics.

Here, we demonstrate a use case of automatic, deep learning-based segmentation of colon epithelium with no requirements for computer programming. We further publish the resulting near pixel accurate dataset of epithelium segmentation of 140 HE stained WSIs and 111 CD3 immunostained WSIs from colon biopsies of both healthy controls and patients with active inflammatory bowel disease.

## 2    Materials and Methods

### 2.1    Dataset of endoscopic colon biopsies

Formalin fixed paraffin embedded (FFPE) biopsies of colonic mucosa were extracted from the NTNU/St. Olavs hospital, Trondheim University Hospital (Norway) biobank of patients with confirmed inflammatory bowel disease or healthy controls with gastrointestinal symptoms but no macroscopic- or microscopic disease. Inclusion and colonoscopies were performed at the Department of Gastroenterology and Hepatology at St. Olavs hospital, Trondheim University Hospital from 2007 to 2018. All patients gave written informed consent and ethical approvals were obtained from the Central Norway Regional Committee for Medical and Health Research Ethics (reference number 2013/212/REKMidt). Consent to publish the anonymized WSI dataset was given by REKMidt in 2021. Each database ID-number used in this study was changed to new anonymized IDs only containing the information "active" or "inactive" disease and whether the WSI has "HE" or "CD3" staining. The full dataset of 251 WSIs with ~31k epithelium annotations is made openly available at https://github.com/andreped/NoCodeSeg with a link to the data repository.

FFPE sections of 4 μm were cut, mounted on slides and either stained with hematoxylin (Mayer's) and Eosin (Y) (HE) or subjected to standard pre-treatment with quenching of endogenous peroxidase and boiling in Tris EDTA pH9 for antigen retrieval before immunohistochemistry. Primary antibody for the T lymphocyte marker was mouse anti-human CD3 (M7254, clone F7.2.38, Dako Agilent, CA, USA), diluted 1:50 in antibody diluent Tris buffer with 0.025% Tween-20 and 1% BSA and incubated overnight at 4 °C. Immunoreactions were visualized with the secondary antibody rabbit/mouse EnVision-HRP/DAB+ kit (K5007, Dako Agilent) and counterstaining with haematoxylin. Omission of the primary antibody was used as negative control and sections from human peripheral lymph node as positive control.

## 2.2    U-Net based epithelial segmentation using QuPath and DeepMIB

The HE and CD3 immunostained slides were scanned using a Hamamatsu NanoZoomer S360 (Hamamatsu Photonics, Japan) scanner at x40 magnification. Slides were imported into the open-source image analysis software QuPath (Bankhead et al., 2017). Epithelium was annotated for ~30 out of 111 CD3 stained WSIs by an experienced gastrointestinal pathologist and checked and corrected by a second pathologist. To make the images compatible with efficient training of semantic segmentation neural networks in DeepMIB, 2048 x 2048 pixels image tiles were exported as 4x downsampled files (from 2048 x 2048 pixels with 512 pixels overlap to a downsampled size of 512 x 512 pixels with 128 pixels overlap) with corresponding binary mask labels (*.png). from QuPath. Overlapping tiles were used to avoid inference errors at the edges of the patches when importing labels back into QuPath. For each WSI, patches containing more than 25% tissue were exported from QuPath.

Images and labels were split randomly into an 80/20% train/test split at the WSI level, such that only unseen WSIs were present in the test set. The data was then placed in separate train and test folders, each containing separate 'Images' and 'Labels' sub-folders.

Two semantic segmentation neural networks were used in this paper: U-Net (Ronneberger et al., 2015) and SegNet (Badrinarayanan et al., 2017). U-Net is a fully-convolutional encoder-decoder neural network initially developed for the purpose of biomedical image segmentation. U-Net is one of two available 2D semantic segmentation networks in DeepMIB which allows optimization of hyperparameters such as U-Net depth, number of filters and input patch size for each segmentation task (Belevich & Jokitalo, 2021). SegNet is a fully-convolutional encoder-decoder neural network where the encoder part is identical to the 13 convolutional layers in the much-used VGG16 network (Badrinarayanan et al., 2017).

A SegNet network with depth of 6 layers with 32 initial filter and input patch size of 256 x 256 pixels was trained until validation loss stagnation around 5% in DeepMIB (MATLAB version 2021a, MIB version 2.8, CUDA version 11.3). The trained SegNet was then used to predict the remaining ~70 WSIs by exporting 4x downsampled 512 x 512 image patches with 128 pixels overlap from QuPath. Patches containing less than 25% tissue were deleted. The resultant images with predicted label files were then loaded in DeepMIB for evaluation and the label patches saved as TIF files. The TIF files were then imported back into QuPath as annotations. Annotations were then confirmed, and errors were manually corrected in QuPath by a pathologist for the remaining WSIs to achieve a dataset of 111 WSIs. A final refinement of the dataset was done by predicting the full dataset and correcting in DeepMIB. The ~5% patches with the lowest mean IoU scores as evaluated inside DeepMIB were exported as text-file lists and the patches could then be copied to a different folder using a Windows PowerShell script (all scripts used in this paper is made available in the NoCodeSeg GitHub repository). The worst performing image patches and their corresponding labels were then loaded and corrected in DeepMIB. A similar strategy was applied to the HE-stained dataset of 140 WSIs, using the U-Net trained on CD3 immunostained WSIs to predict and correct an initial batch of ~30 HE-stained WSIs. Then training a U-Net on the initial batch of annotated HE-stained WSIs, applying it on the remaining HE-stained WSIs, and retraining the U-Net. The final datasets (140 HE-stained and 111 CD3 immunostained WSIs, or 6322 HE and 4323 CD3 4x downsampled 512 x 512 image patches in each dataset) were again split into an 80/20% train/test split at the WSI level, such that 36 (HE) and 21 (CD3) previously unseen WSIs were present in the test set and new networks were trained from scratch using DeepMIB to assess the performance of the software on this larger train/test set (see Table 1).

Finally, two CNNs, SegNet and U-Net, were then trained using DeepMIB. To achieve maximum variety of different image patches per mini batch, the number of patches DeepMIB extracts per image in a single mini batch was set to one. Initially the number of patches per image was set to the same number as the number of applied augmentations, however, this produced inferior results to using just one patch per image per mini batch. Three percent of the training set images were randomly chosen by DeepMIB for the validation set. A fixed random generator seed was used to make comparison between training different conditions more direct. Several hyperparameters were tested, such as variable input patch size (128 x 128, 256 x 256, 512 x 512), number of filters (16, 24, 32, 64), network depth (4, 5, 6, 7, 8), and the presence and absence of augmentations. Finally, U-Net and SegNet were trained for 200 epochs, which was the number of epochs required for training loss stagnation. Further global training settings were as follows: Padding: Same; Solver: Adam; Shuffle: Every-epoch; Initial learning rate: 0.00005; L2 Regularization: 0.0001; Decay rate of gradient moving average: 0.9; Decay rate of squared gradient moving average: 0.999. Augmentations used in all described trainings were performed in a blended fashion (MIB version 2.8) with a 30% probability for each augmentation to be applied to each augmented image patch during training. The fraction of images for augmentation was set to 75%, i.e. 25% of input image patches were not augmented, while 75% had a 30% chance of being augmented with either of the following augmentations [numeric limits show in brackets]: Random left-right/top-bottom reflections, random 90/270-degree rotations, random X/Y/X+Y image scaling [1.0, 1.1], random color augmentation: Hue [-0.03, 0.03], saturation [-0.05, 0.05], random intensity augmentation: brightness [-0.1, 0.1], contrast [0.9, 1.1], and zero-mean Gaussian blur with standard deviation in range [0, 0.5].

A selection of metrics was extracted for each patch from DeepMIB, and metrics were then averaged at the WSI-level (Table 1). The reported metrics were produced from calculating the WSI-level average. The following metrics were calculated: micro and macro-averaged pixel-wise accuracy, macro and weighted IoU, and class-wise DSC for the exterior Epithelium classes. U-Net proved to consistently outperform SegNet for both the HE and CD3 dataset (see Table 1). Initially, increasing input patch sizes were tested, with size 64 x 64, 128 x 128, 256 x 256, and 512 x 512. The available 24 GB GPU allowed a maximum batch size of 16 for a U-Net with 512 x 512 patch size, 32 filter, and depth of 6. Thus, these settings (16 batch size, 32 filter, depth 6) were kept for all the different input patch size trainings to be comparable.

### 2.3 Deployment in FastPathology

The best performing trained U-Net model from DeepMIB was exported to the ONNX format using the ExportONNXNetwork method from the Deep Learning Toolbox in MATLAB. As ONNX does not currently support MATLAB's implementation of the UnPooling operation in SegNet, U-Net was the only model converted to ONNX.

We defined an inference pipeline consisting of applying the trained segmentation model across the WSI in an overlapping, sliding window fashion, similarly as done in a previous study (Pedersen et al., 2021). The result of each patch was binarized using a threshold of 0.5, before being stitched to form a tiled, pyramidal image. When inference was complete, the resulting pyramidal image was exported to the disk in the open TIFF format.

To demonstrate the performance of FastPathology, runtime experiments were conducted. Runtimes were measured for the total inference pipeline, as well as for individual pipeline components (runtimes reported are without overlapping inference). The experiments were repeated ten times for the same WSI, using three different inference engines (OpenVINO CPU, OpenVINO GPU, and TensorRT). For each metric, the average of the ten runs were reported. The source code to reproduce the experiments can be found on GitHub[7].

## 2.4    Computer hardware

Runtime experiments were performed on a Razer Blade 15 Base laptop, with an Intel i7-10750H CPU @ 2.60 GHz, 32 GB RAM, an Intel UHD graphics integrated GPU, and NVIDIA RTX 2070 Max-Q (8 GB) dedicated GPU. All other analyses were performed on a Dell Precision 5820 Tower, with an Intel(R) Xeon(R) W-2155 CPU @ 3.30GHz, 96 GB RAM, and a NVIDIA Titan RTX (24 GB) dedicated GPU.

## 3    Results

### 3.1    U-Net based epithelial segmentation using QuPath and DeepMIB

An experienced gastro pathologist annotated epithelium for 30 WSIs using QuPath, an efficient manual annotation software for large gigapixel WSIs. This was checked and corrected by a second pathologist. Image patches were then exported from QuPath with corresponding masks. WSIs often contain 50-90% white background, which will make the exterior class completely dominant in training. Therefore, a glass detection method was used, similarly as done in a previous study (Pedersen et al., 2021), and patches with less than 25% tissue were discarded.

### 3.1.2    Semantic segmentation of colon epithelium using DeepMIB

We performed several trainings in DeepMIB using two different CNNs (i.e., SegNet and U-Net), with a variety of hyperparameters to find the highest performance (see Materials and Methods, section 2.2). Initially a SegNet network was trained and applied to new unannotated WSIs. Annotations were imported and manually corrected in QuPath by a pathologist. Subsequent training cycles were performed with U-Net 512x512 in a repetitive fashion described in Figure 1 (DeepMIB training, inference of new WSIs and import into QuPath for correction of annotations, export for new DeepMIB training, etc.) to achieve a final dataset of 111 WSIs (see Figure 1). A final refinement of the annotations was done by exporting individual accuracy scores for all image patches exported from DeepMIB. This allowed sorting of the patches which was in most disagreement with the U-Net predictions (typically mean IoU scores below 0.85). The set of ~5% worst performing patches were then loaded in DeepMIB such that a pathologist could refine the annotations directly on several hundred image patches instead of going through the whole dataset of ~5000-7000 image patches. This final refinement made it possible to achieve almost pixel accurate epithelial segmentation of approximately 100 WSIs. The top-performing CD3-trained network was used to repeatedly predict and correct the 140 HE stained WSIs, following the workflow described in Figure 1.

---

[7] https://github.com/andreped/NoCodeSeg

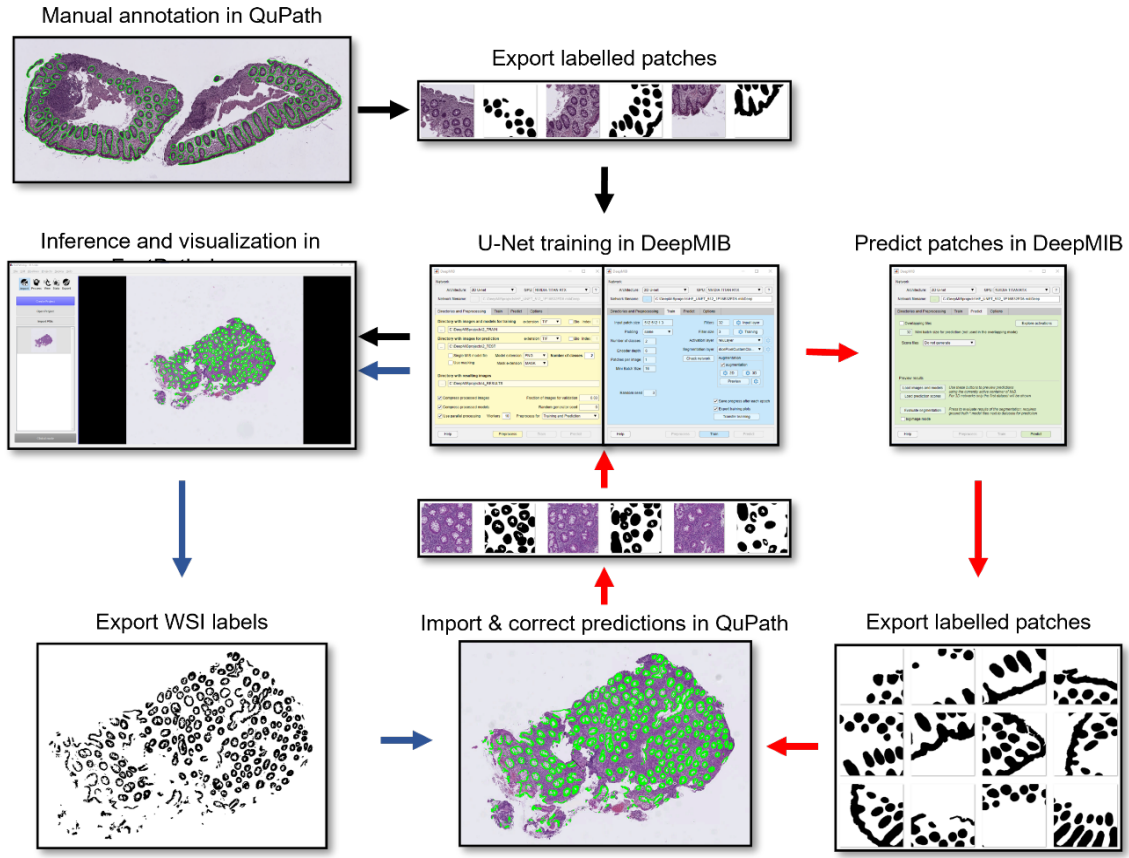

Figure *1*: Flowchart showing the pipeline from manual annotation in QuPath, export of labelled patches from QuPath, CNN training in DeepMIB, expansion of the dataset by predicting unseen WSIs in DeepMIB and importing and correcting predictions in QuPath, and final export of trained networks as ONNX-files and rapid prediction directly on WSIs in FastPathology.

This resulted in a final dataset of fully annotated patches, from 140 HE stained and 111 CD3 immunostained colon biopsy WSIs (see Figure 2 for examples). The datasets were split into the two subsets: train (80%; n=104 HE; n=90 CD3) and test set (20%; n=36 HE; n=21 CD3). Two segmentation networks (i.e., U-Net and SegNet) were then trained on the final refined datasets to assess performance. We limited each training to 200 epochs and a similar global training setting (see Materials and Methods) for comparable results (see Table 1).

Table 1: Comparative accuracies on the HE stained (n=38) and CD3 immunostained (n=21) test sets with different hyperparameter settings for U-Net and SegNet.

| STAIN | ARCH. | PATCH SIZE | NR. OF FILT. | DEPTH | BATCH SIZE | MICRO ACC | MACRO ACC | MACRO IOU | WEIG. IOU | EXT. DSC | EPITH. DSC |
|---|---|---|---|---|---|---|---|---|---|---|---|
| HE | U-Net | 512 x 512 | 32 | 6 | 16 | **0.989** | 0.972 | **0.955** | **0.978** | **0.992** | **0.953** |
| HE | U-Net | 256 x 256 | 32 | 6 | 16 | **0.989** | **0.983** | 0.938 | **0.978** | **0.992** | 0.920 |
| HE | U-Net | 256 x 256 | 32 | 6 | 32 | 0.988 | 0.978 | 0.936 | 0.976 | 0.991 | 0.920 |
| HE | U-Net | 256 x 256 | 64 | 6 | 32 | 0.987 | 0.974 | 0.935 | 0.975 | 0.991 | 0.919 |
| HE | U-Net | 128 x 128 | 32 | 6 | 16 | 0.988 | **0.983** | 0.932 | 0.977 | 0.991 | 0.911 |
| HE | U-Net | 64 x 64 | 32 | 6 | 16 | 0.985 | 0.965 | 0.924 | 0.971 | 0.989 | 0.904 |
| HE | SegNet | 512 x 512 | 32 | 6 | 16 | 0.983 | 0.964 | 0.928 | 0.967 | 0.988 | 0.918 |
| HE | SegNet | 256 x 256 | 32 | 6 | 16 | 0.987 | 0.973 | 0.939 | 0.974 | 0.991 | 0.927 |
| HE | SegNet | 128 x 128 | 32 | 6 | 16 | 0.979 | 0.964 | 0.904 | 0.960 | 0.985 | 0.884 |
| CD3 | U-Net | 512 x 512 | 32 | 6 | 16 | **0.990** | **0.981** | **0.955** | **0.980** | **0.992** | **0.948** |
| CD3 | U-Net | 256 x 256 | 32 | 6 | 16 | 0.987 | 0.977 | 0.931 | 0.974 | 0.990 | 0.911 |
| CD3 | SegNet | 512 x 512 | 32 | 6 | 16 | 0.976 | 0.953 | 0.920 | 0.954 | 0.983 | 0.919 |
| CD3 | SegNet | 256 x 256 | 32 | 6 | 16 | 0.971 | 0.949 | 0.898 | 0.945 | 0.979 | 0.889 |

*All metrics were reported as the mean at WSI-level. Best performing methods are highlighted in bold, for each respective metric and for each data set. ARCH: Architecture, FILT: Filters, NR: Number, ACC: Accuracy, WEIG: Weighted, EXT: Exterior, IOU: Intersection over Union, DSC: Dice Similarity Coefficient, EPITH: Epithelial, HE: Hematoxylin-Eosin, CD3: T-cell lymphocyte immunomarker.*

For the U-Net models on the HE dataset, an increase in segmentation accuracy was observed with increasing input patch sizes from 64 x 64 (Epithelium DSC 0.904) to 512 x 512 (Epithelium DSC 0.953). The best segmentation accuracy for SegNet was observed with input patch size 256 x 256 for the HE dataset (Epithelium DSC 0.927). For the CD3 dataset, the maximum segmentation accuracy was observed for 512 x 512 input patches for both U-Net and SegNet (Epithelium DSC of 0.948 and 0.919, respectively).

There was a negligible difference in segmentation accuracy when increasing the number of filters from 32 to 64 (Epithelium DSC 0.920 vs. 0.919) or increasing the batch size from 16 to 32 (Epithelium DSC 0.920 vs. 0.920). U-Net consistently outperformed SegNet with a top segmentation accuracy of Epithelium DSC of 0.953 and 0.948 (HE and CD3, U-Net 512 x 512) vs. 0.927 and 0.919 (HE SegNet 256 x 256 and CD3 SegNet 512 x 512). Further testing with different depth of the networks was also performed, but depth 6 seemed to perform consistently higher (data not shown).

Using our best performing 256 x 256 U-Net model, the proposed inference pipeline took ~5.60 seconds to complete for the entire WSI using FastPathology (see Table 2). In our experience, this is well within the range for running direct inference in a clinical setting, and even the longest CPU-based inference times would probably not be limiting to the use of such algorithms by pathologists. The fastest inference engine was TensorRT, whereas using

OpenVINO took ~76.5 seconds (a 13.7x improvement using TensorRT). The main bottleneck of the pipeline was the neural network inference. For OpenVINO, ~94.3 % of the patch runtime was due to inference alone, whereas for TensorRT this was only ~49.5 %. Using TensorRT, our inference pipeline required ~2.1 GB of VRAM and ~4.2 GB of RAM for running inference on a full WSI with a network trained with patch sizes of 256 x 256 pixels.

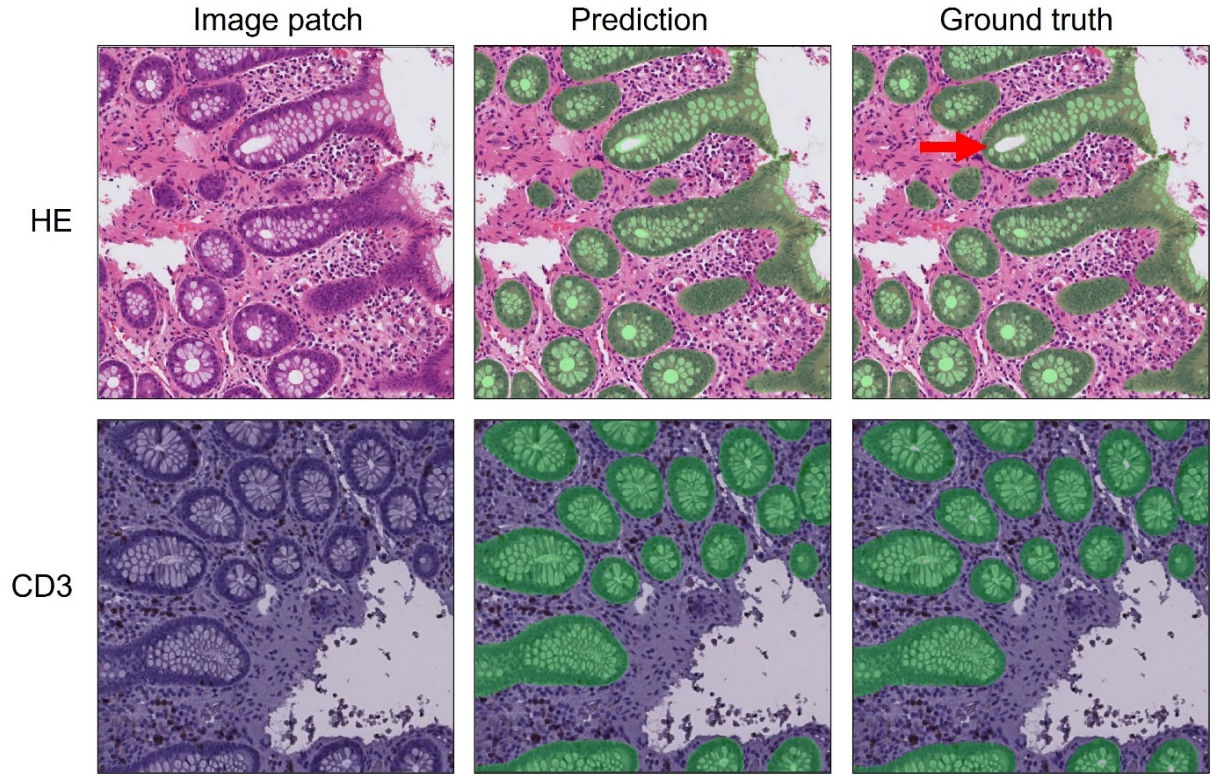

Figure 2: Examples of predictions (middle column) and ground truth (right column) of epithelial segmentation (transparent green) of HE stained (top row) and CD3 immunostained (bottom row) 512 x 512-pixel image patches in DeepMIB. The arrow shows the approximate cut-offs for (filled or unfilled) minimal tubule hole size used during annotation.

Table 2: Runtime measurements of different inference engines using FastPathology.

| INFERENCE ENGINE | PROCESSOR | PATCH GEN. (MS) | NN INPUT (MS) | NN INFERENCE (MS) | NN OUTPUT (MS) | PATCH STITCHER (MS) | EXPORT WSI TIFF (MS) | TOTAL TIME/WSI (S) |
|---|---|---|---|---|---|---|---|---|
| OPENVINO CPU | Intel i7-10750H | 3.65 | 1.03 | 135.31 | 0.80 | 2.76 | 7.09 | 76.38 |
| OPENVINO GPU | Intel UHD graphics | **3.29** | 1.26 | 133.96 | 1.25 | 3.46 | 7.83 | 76.65 |
| TENSORRT | RTX 2070 Max-Q | 5.12 | **0.80** | **7.31** | **0.19** | **1.35** | **5.40** | **5.60** |

*The table shows means of 10 runtime experiments for the 256x256 pixel input patch size U-Net applied to a single WSI (540 patches). Inference measurements show runtime per 256x256 patch in milliseconds (ms). Export of a full WSI pyramidal TIFF performed once after inference is reported in ms, and the total runtime for the full WSI (including TIFF*

*export) is shown in seconds (s). The fastest runtimes are highlighted in bold. GEN: Generator, NN: Neural Network, WSI: Whole Slide Image.*

## 5　　Discussion

### 5.1　　Benefits and limitations with using multiple software solutions

The motivation of this study was to segment epithelium in a large dataset from a biobank of normal and diseased (inflammatory bowel disease) colon biopsies. We aimed to achieve this without the need for computer coding abilities, while simultaneously taking advantage of the strongest sides of available open-source software solutions. We demonstrate an open-source, free-to-use pipeline that can achieve high accuracy segmentation of histopathological WSIs available to a broad user base without the ability to write computer code. We further demonstrate the advantages of using open-source, non-proprietary software and formats that can be exchanged between these three software packages. The pipeline could be improved by incorporating all tasks into a single software solution. However, the use of several software solutions and exchange of information between them makes it possible to use more specialized open-source solutions best suited for each task – QuPath for annotations of whole slide images, DeepMIB for neural network training, and FastPathology for efficient inference and visualization of trained models. A disadvantage of such a multi-software pipeline is that it requires three separate software installations, which over time might diverge in compatibility and use different versions of auxiliary software, such as versions of CUDA.

Even though the pipeline does not require the ability to write computer code, it does require the use of some scripts, such as the QuPath export/import scripts, which requires copy/pasting of pre-existing code, and perhaps also changing of some parameters within those scripts to make the pipeline suitable for different tasks. In the near future, it is likely that this will be possible solely through the GUI in QuPath.

The epithelial segmentation accuracy was comparably high for both the best performing U-Nets on HE (DSC Epithelium 0.953) and CD3 images (DSC Epithelium 0.948), demonstrating the robustness of U-Net for this task. Segmentation accuracy was generally better with larger patch sizes (512 x 512 vs. 256 x 256 DSC Epithelium 0.953 vs. 0.920), however 256 x 256 patch size networks require much less GPU memory for training and inference. We have not compared the segmentation accuracy of our trained models to current state-of-the-art architectures (Tao et al., 2020; Yuan et al., 2019). However, DSC scores for the epithelial class up to ~95% on unseen test sets show little room for considerable improvement, making the U-Net segmentation accuracy for these data sets probably near state-of-the-art. It has also been argued by others (Isensee et al., 2021), that there is little to gain from changing neural network architecture for semantic segmentation. The U-Net architecture presented can also easily be tuned code-free to be better suited for a specific task. The datasets are published with this paper and comparison to state-of-the-art models will therefore be possible by others.

### 5.3　　The dataset and annotations

Several issues arose during annotation. Defining a pixel-accurate epithelium ground truth is difficult as several images contain artifacts (folds, blurred areas, poorly fixated tissue, stain exudates, etc.) as well as intraepithelial inclusions (e.g., granulocytes) (see Figure 3). These cannot be easily defined into the dichotomous categories: epithelium or exterior, as e.g., folded tissue might contain both classes. Therefore, slides with more than ~10-20% artifacts were excluded from the dataset, as they contained large areas not suitable for pathological diagnostics either. Furthermore, defining intraepithelial granulocytes as part of the epithelium or not had to be individually considered, as large abscess like assemblies of granulocytes with little or no visible epithelium can obviously not be considered epithelium.

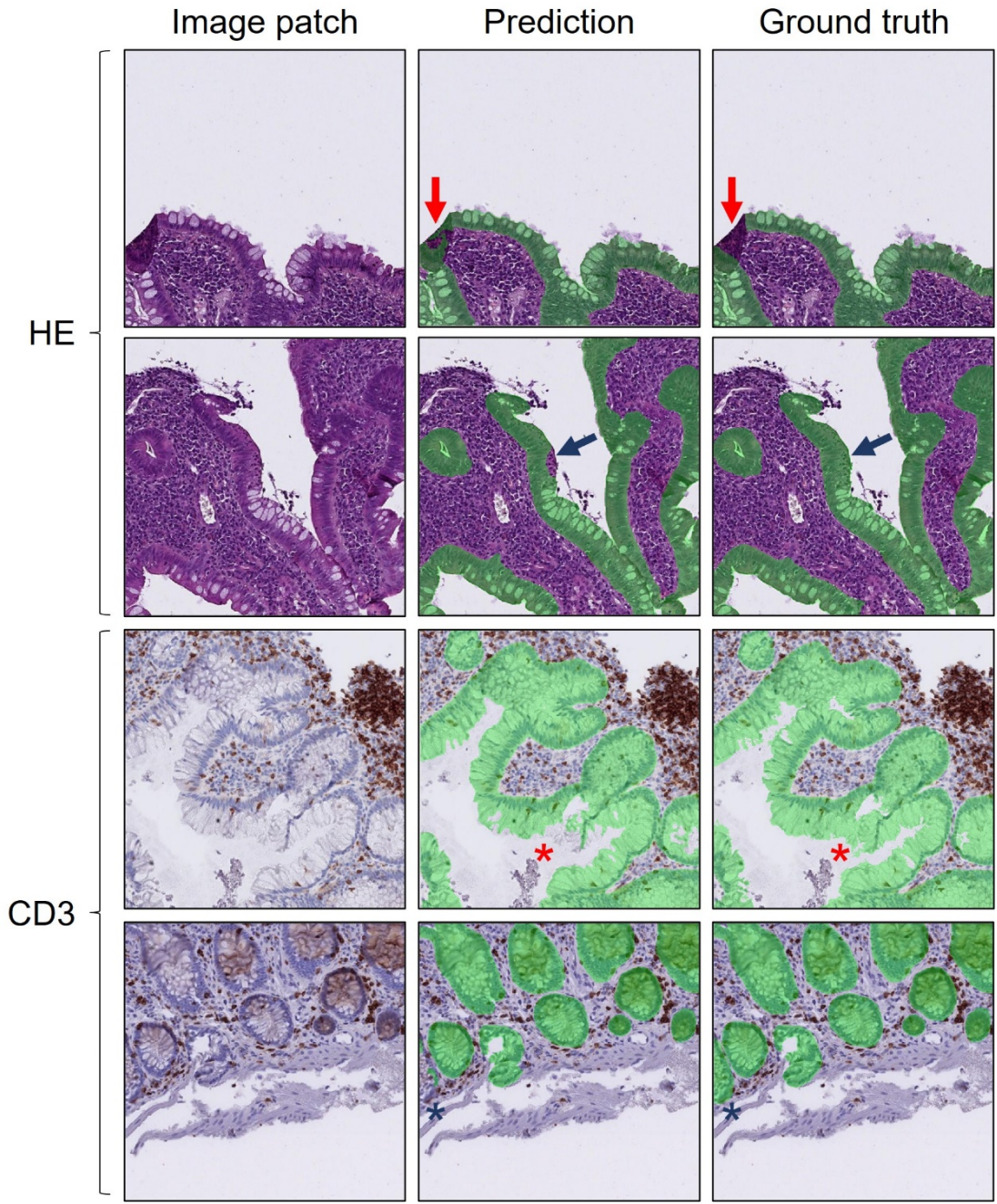

Figure 3: Examples of prediction errors in difficult regions: HE stained images with folding artifacts (top row, red arrows) and granulocyte aggregates (second row, blue arrows). CD3 immunostained images with thick mucin rich epithelium (third row, red stars) and poorly fixated blurred epithelium at the edge of a patch (bottom row, blue stars). Prediction (middle

column) and ground truth (right column) of epithelial segmentation are shown in transparent green. 512 x 512-pixel image patches displayed in DeepMIB.

However, the clinical use of an epithelium segmentation algorithm in colon biopsies would certainly involve quantitative estimates of intraepithelial granulocytes and excluding large granulocytic abscesses during annotation also potentially diminishes the clinical value of the algorithm. Indeed, significant differences in prediction accuracies were seen for the test sets of both HE and CD3 immunostained slides between patients with active disease (with infiltration of neutrophilic granulocytes) and inactive disease (see Figure 4). Still, the segmentation accuracy was deemed to be at a clinically acceptable level with Epithelium DSC scores >91% for all slides.

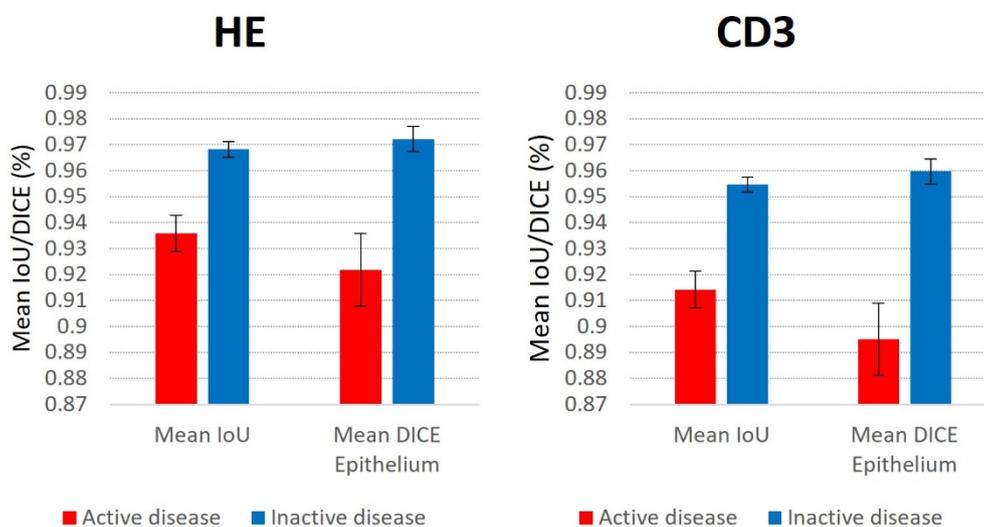

Figure 4: Significant differences in prediction accuracies for the HE stained test set WSIs (n=36) with active disease (n=15) vs. inactive disease (n=21), and CD3 immunostained test set WSIs (n=21) with active disease (n=7) and inactive disease (n=14). Error bars represent 95% confidence intervals assuming normality. Two-tailed Student's T-test of active vs. inactive disease gave p < 0.0001 for all four comparisons.

The cytoplasmic part of colon epithelium has a wide variation in size, particularly because of variation of mucin content. Inconsistencies in the cut-off for when mucin is no longer part of the epithelial cell and starts being part of the exterior class, was an obvious source of deviation between ground truth and predictions (see Figure 3). Furthermore, the cut-off between when the lumen of the colonic tubule ceases to be part of the epithelium and starts being part of the exterior class, was problematic. This was alleviated to a certain extent by taking advantage of the power QuPath has as an annotation tool which allows running a single background thresholder pixel classifier algorithm, subsequently creating several large and small background annotations. These could subsequently be selected by a minimal size cut-off and subtracted from the epithelium annotations consistently for the entire dataset by running QuPath scripts in batch mode. A similar procedure is also possible to perform in DeepMIB using the BWThresholding tool followed by subtraction from all annotations, then a small dilation and subsequent similar erosion to fill small holes. One should be aware that this, however, might introduce merging of nearby annotations. The top row of figure 2 (HE segmentation results) provides a visual approximation of the maximal colonic tubule lumen sizes that are accepted as being part of the epithelium class (transparent green) or exterior class.

## 6      Conclusion

In this paper, we have presented a code-free pipeline for developing and deploying deep neural network segmentation models for computational pathology. The pipeline uses open, free software and enables the user to build and test state-of-the-art deep learning methods for segmentation of WSIs, without requiring any programming experience. We also demonstrate competitive results on two segmentation tasks with rapid inference of about 5 seconds for an entire WSI. The WSIs and annotations are also made publicly available to contribute to the active research within the field.

## 7      Author contributions

HP, AP, IBe, ER, IBa: Writing of initial draft. ER, IBa: Provided data material. ER, IBa: Identified and selected patients from the biobank. ER: Performed the initial annotations and preparation/scanning of the slides. HP: Iteratively improved annotations using the described QuPath/DeepMIB pipeline and performed all DeepMIB semantic segmentation trainings. AP: Performed runtime experiments. AP, ES, IBe: Improved respective software to be better suited for this application. IR/EJ: Supervised the structuring of the paper and code development for FastPathology/DeepMIB. All authors contributed to reviewing and finalizing the manuscript.

## 8      Funding

This work was funded by the Research Fund for the Center for Laboratory Medicine, St. Olavs hospital, Trondheim University Hospital, the Faculty of Medicine and Health Sciences, NTNU and the Liaison Committee between the Central Norway Regional Health Authority and NTNU. IBe and EJ are supported by Biocenter Finland and Academy of Finland (project 1331998). Funding for open access publishing was given by NTNU University Library's Open access publishing fund.

## 9      Acknowledgments

We would like to acknowledge Pete Bankhead (University of Edinburgh), Melvin Gelbard and the rest of the support team working with QuPath for continual feedback and help with sharing scripts and resolving issues with QuPath. The QuPath-related export/import scripts in the GitHub repository are inspired by their shared scripts. We thank the staff of the Gastrointestinal Endoscopy Unit, Department of Gastroenterology and Hepatology, St. Olavs hospital, Trondheim University Hospital for support with sample collection, Bjørn Munkvold at Department of Clinical and Molecular Medicine (IKOM), NTNU for technical assistance and the Department of Pathology at St. Olavs hospital, Trondheim University Hospital for scanning the tissue sections.

## 10     Data Availability Statement

The datasets generated for this study can be found in the GitHub repository (https://github.com/andreped/NoCodeSeg) with link to the data repository.

## References

Abadi, M., Agarwal, A., Barham, P., Brevdo, E., Chen, Z., Citro, C., Corrado, G. S., Davis, A., Dean, J., Devin, M., Ghemawat, S., Goodfellow, I., Harp, A., Irving, G., Isard, M., Jia, Y., Jozefowicz, R., Kaiser, L., Kudlur, M., Levenberg, J., Mane, D., Monga, R., Moore, S., Murray, D., Olah, C., Schuster, M., Shlens, J., Steiner, B., Sutskever, I., Talwar, K., Tucker, P., Vanhoucke, V., Vasudevan, V., Viegas, F., Vinyals, O., Warden, P., Wattenberg, M., Wicke, M., Yu, Y., & Zheng, X. (2016). TensorFlow: Large-Scale Machine Learning on Heterogeneous Distributed Systems. arXiv:1603.04467. Retrieved March 01, 2016, from https://ui.adsabs.harvard.edu/abs/2016arXiv160304467A

Badrinarayanan, V., Kendall, A., & Cipolla, R. (2017). SegNet: A Deep Convolutional Encoder-Decoder Architecture for Image Segmentation. *Ieee Transactions on Pattern Analysis and Machine Intelligence*, *39*(12), 2481-2495. https://doi.org/10.1109/Tpami.2016.2644615

Bankhead, P., Loughrey, M. B., Fernandez, J. A., Dombrowski, Y., McArt, D. G., Dunne, P. D., McQuaid, S., Gray, R. T., Murray, L. J., Coleman, H. G., James, J. A., Salto-Tellez, M., & Hamilton, P. W. (2017). QuPath: Open source software for digital pathology image analysis. *Sci Rep*, *7*(1), 16878. https://doi.org/10.1038/s41598-017-17204-5

Belevich, I., Joensuu, M., Kumar, D., Vihinen, H., & Jokitalo, E. (2016). Microscopy Image Browser: A Platform for Segmentation and Analysis of Multidimensional Datasets. *PLoS Biol*, *14*(1), e1002340. https://doi.org/10.1371/journal.pbio.1002340

Belevich, I., & Jokitalo, E. (2021). DeepMIB: User-friendly and open-source software for training of deep learning network for biological image segmentation. *PLoS Comput Biol*, *17*(3), e1008374. https://doi.org/10.1371/journal.pcbi.1008374

Djuric, U., Zadeh, G., Aldape, K., & Diamandis, P. (2017). Precision histology: how deep learning is poised to revitalize histomorphology for personalized cancer care. *NPJ Precis Oncol*, *1*(1), 22. https://doi.org/10.1038/s41698-017-0022-1

Falk, T., Mai, D., Bensch, R., Cicek, O., Abdulkadir, A., Marrakchi, Y., Bohm, A., Deubner, J., Jackel, Z., Seiwald, K., Dovzhenko, A., Tietz, O., Dal Bosco, C., Walsh, S., Saltukoglu, D., Tay, T. L., Prinz, M., Palme, K., Simons, M., Diester, I., Brox, T., & Ronneberger, O. (2019). U-Net: deep learning for cell counting, detection, and morphometry (vol 16, pg 67, 2019). *Nature Methods*, *16*(4), 351-351. https://doi.org/10.1038/s41592-019-0356-4

Hu, M. D., & Edelblum, K. L. (2017). Sentinels at the frontline: the role of intraepithelial lymphocytes in inflammatory bowel disease. *Curr Pharmacol Rep*, *3*(6), 321-334. https://doi.org/10.1007/s40495-017-0105-2

Isensee, F., Jaeger, P. F., Kohl, S. A. A., Petersen, J., & Maier-Hein, K. H. (2021). nnU-Net: a self-configuring method for deep learning-based biomedical image segmentation. *Nat Methods*, *18*(2), 203-211. https://doi.org/10.1038/s41592-020-01008-z

Jahn, S. W., Plass, M., & Moinfar, F. (2020). Digital Pathology: Advantages, Limitations and Emerging Perspectives. *Journal of Clinical Medicine*, *9*(11), 3697. https://doi.org/10.3390/jcm9113697

Lara, H., Li, Z. B., Abels, E., Aeffner, F., Bui, M. M., ElGabry, E. A., Kozlowski, C., Montalto, M. C., Parwani, A. V., Zarella, M. D., Bowman, D., Rimm, D., & Pantanowitz, L. (2021). Quantitative Image Analysis for Tissue Biomarker Use: A White Paper From the Digital Pathology Association. *Applied Immunohistochemistry & Molecular Morphology*, *29*(7), 479-493. https://doi.org/10.1097/Pai.0000000000000930

Lutnick, B., Ginley, B., Govind, D., McGarry, S. D., LaViolette, P. S., Yacoub, R., Jain, S., Tomaszewski, J. E., Jen, K. Y., & Sarder, P. (2019). An integrated iterative annotation technique for easing neural network training in medical image analysis. *Nat Mach Intell*, *1*(2), 112-119. https://doi.org/10.1038/s42256-019-0018-3

Lutter, L., Hoytema van Konijnenburg, D. P., Brand, E. C., Oldenburg, B., & van Wijk, F. (2018). The elusive case of human intraepithelial T cells in gut homeostasis and inflammation. *Nat Rev Gastroenterol Hepatol*, *15*(10), 637-649. https://doi.org/10.1038/s41575-018-0039-0

Miao, R., Toth, R., Zhou, Y., Madabhushi, A., & Janowczyk, A. (2021). Quick Annotator: an open-source digital pathology based rapid image annotation tool. arXiv:2101.02183. Retrieved January 01, 2021, from https://ui.adsabs.harvard.edu/abs/2021arXiv210102183M

Moen, E., Bannon, D., Kudo, T., Graf, W., Covert, M., & Van Valen, D. (2019). Deep learning for cellular image analysis. *Nat Methods*, *16*(12), 1233-1246. https://doi.org/10.1038/s41592-019-0403-1

Olivares-Villagomez, D., & Van Kaer, L. (2018). Intestinal Intraepithelial Lymphocytes: Sentinels of the Mucosal Barrier. *Trends Immunol*, *39*(4), 264-275. https://doi.org/10.1016/j.it.2017.11.003

Pedersen, A., Valla, M., Bofin, A. M., De Frutos, J. P., Reinertsen, I., & Smistad, E. (2021). FastPathology: An Open-Source Platform for Deep Learning-Based Research and Decision Support in Digital Pathology. *Ieee Access*, *9*, 58216-58229. https://doi.org/10.1109/Access.2021.3072231

Ronneberger, O., Fischer, P., & Brox, T. (2015). U-Net: Convolutional Networks for Biomedical Image Segmentation. *Medical Image Computing and Computer-Assisted Intervention, Pt Iii*, *9351*, 234-241. https://doi.org/10.1007/978-3-319-24574-4_28

Schmidt, U., Weigert, M., Broaddus, C., & Myers, G. (2018). Cell Detection with Star-Convex Polygons. *Medical Image Computing and Computer Assisted Intervention - Miccai 2018, Pt Ii*, *11071*, 265-273. https://doi.org/10.1007/978-3-030-00934-2_30

Smistad, E., Bozorgi, M., & Lindseth, F. (2015). FAST: framework for heterogeneous medical image computing and visualization. *International Journal of Computer Assisted Radiology and Surgery*, *10*(11), 1811-1822. https://doi.org/10.1007/s11548-015-1158-5

Smistad, E., Ostvik, A., & Pedersen, A. (2019). High Performance Neural Network Inference, Streaming, and Visualization of Medical Images Using FAST. *Ieee Access*, *7*, 136310-136321. https://doi.org/10.1109/Access.2019.2942441

Srinidhi, C. L., Ciga, O., & Martel, A. L. (2021). Deep neural network models for computational histopathology: A survey. *Med Image Anal*, *67*, 101813. https://doi.org/10.1016/j.media.2020.101813

Tao, A., Sapra, K., & Catanzaro, B. (2020). Hierarchical Multi-Scale Attention for Semantic Segmentation. arXiv:2005.10821. Retrieved May 01, 2020, from https://ui.adsabs.harvard.edu/abs/2020arXiv200510821T

van der Laak, J., Litjens, G., & Ciompi, F. (2021). Deep learning in histopathology: the path to the clinic. *Nat Med*, *27*(5), 775-784. https://doi.org/10.1038/s41591-021-01343-4

von Chamier, L., Laine, R. F., Jukkala, J., Spahn, C., Krentzel, D., Nehme, E., Lerche, M., Hernandez-Perez, S., Mattila, P. K., Karinou, E., Holden, S., Solak, A. C., Krull, A., Buchholz, T. O., Jones, M. L., Royer, L. A., Leterrier, C., Shechtman, Y., Jug, F., Heilemann, M., Jacquemet, G., & Henriques, R. (2021). Democratising deep learning for microscopy with ZeroCostDL4Mic. *Nat Commun*, *12*(1), 2276. https://doi.org/10.1038/s41467-021-22518-0

Yuan, Y., Chen, X., Chen, X., & Wang, J. (2019). Segmentation Transformer: Object-Contextual Representations for Semantic Segmentation. arXiv:1909.11065. Retrieved September 01, 2019, from https://ui.adsabs.harvard.edu/abs/2019arXiv190911065Y